\documentclass[journal]{IEEEtran}
\ifCLASSINFOpdf
   \usepackage[pdftex]{graphicx}
\else
\fi
\usepackage{amsmath}
\usepackage{hyperref}
\usepackage{url}

\usepackage{cite}

\usepackage{multirow}
\usepackage{multicol}
\usepackage{mathtools}
\usepackage{bm}

\usepackage{amsmath,epsfig}
\usepackage{amsfonts}
\usepackage{url}
\usepackage{amssymb}
\usepackage{amsthm}

\usepackage{caption}
\usepackage{subcaption}

\begin{document}
%
\title{Onboard deep lossless and near-lossless predictive coding of hyperspectral images with line-based attention}
%
%
%

\author{Diego~Valsesia, \IEEEmembership{Member, IEEE},
        Tiziano~Bianchi, \IEEEmembership{Member, IEEE}
        and~Enrico~Magli, \IEEEmembership{Fellow, IEEE}%
        \thanks{The authors are with Politecnico di Torino -- Department of Electronics and Telecommunications, Italy. email: \{name.surname\}@polito.it. This study was carried out within the FAIR - Future Artificial Intelligence Research and received funding from the European Union Next-GenerationEU (PIANO NAZIONALE DI RIPRESA E RESILIENZA (PNRR) – MISSIONE 4 COMPONENTE 2, INVESTIMENTO 1.3 – D.D. 1555 11/10/2022, PE00000013). This manuscript reflects only the authors’ views and opinions, neither the European Union nor the European Commission can be considered responsible for them.}
}

\maketitle

\begin{abstract}
Deep learning methods have traditionally been difficult to apply to compression of hyperspectral images onboard of spacecrafts, due to the large computational complexity needed to achieve adequate representational power, as well as the lack of suitable datasets for training and testing. In this paper, we depart from the traditional autoencoder approach and we design a predictive neural network, called LineRWKV, that works recursively line-by-line to limit memory consumption. In order to achieve that, we adopt a novel hybrid attentive-recursive operation that combines the representational advantages of Transformers with the linear complexity and recursive implementation of recurrent neural networks. The compression algorithm performs prediction of each pixel using LineRWKV, followed by entropy coding of the residual. Experiments on the HySpecNet-11k dataset and PRISMA images show that LineRWKV is the first deep-learning method to outperform CCSDS-123.0-B-2 at lossless and near-lossless compression. Promising throughput results are also evaluated on a 7W embedded system.
\end{abstract}

\begin{IEEEkeywords}
Hyperspectral image compression, deep learning, self-attention, RWKV, predictive coding
\end{IEEEkeywords}

%
\IEEEpeerreviewmaketitle

\section{Introduction}
\label{sec:intro}

Hyperspectral images acquired by spacecrafts are an essential tool for Earth observation, playing pivotal roles in environmental monitoring, urban planning, tackling climate change and much more. The spatial and spectral resolution of instruments keeps growing to satisfy the demands of final users. However, this poses significant challenges for the transmission and management of huge amounts of spatial and spectral information, making compression a topic of paramount importance. Even modest gains in rate-distortion performance may result in significant bandwidth reduction, leading to higher image availability or even better resolutions. 

The challenge of onboard compression of hyperspectral images is far from trivial, as it requires methods that are efficient in processing massive amounts of data with the limited computational resources that are available on a satellite, and, at the same time, are able to capture complex spatial-spectral redundancy patterns. In the last few years, the community has settled around the CCSDS-123.0-B-2 standard \cite{ccsds123,cabronero2021ccsds} which exploits predictive coding to achieve an excellent tradeoff between rate-distortion performance and complexity.
Meanwhile, deep learning has revolutionized approaches in various fields, leading to new and more powerful models for a huge variety of tasks, including image and video compression. It is then only natural to wonder if onboard compression of hyperspectral images can benefit from such techniques. 
However, the answer is far from obvious. On one hand, traditional methods have worked very well; the CCSDS-123.0-B-2 recommendation employs a simple adaptive predictor which learns from past samples; instead, adaptivity is difficult to achieve using neural networks, and requires Transformers-like architectures that provide a form of context dependence. On the other hand, it is well-known that non-linear models (even simple ones as employed in the JPEG-LS standard \cite{jpegls}) are very good at capturing image detail and discontinuities; the CCSDS-123.0-B-2 recommendation employs a fully linear model, whereas neural networks are a perfect fit for learning nonlinear functions. 
Moreover, effective compression of hyperspectral images with neural networks requires careful modeling of 3D spatial-spectral features, which is even more challenging in presence of significant memory and computational constraints. Some approaches have been studied to extend techniques for RGB images, largely based on autoencoders, to the hyperspectral case while constraining complexity \cite{alves2021reduced,chong2021end,dua2021convolution,la2022hyperspectral,mijares2023scalable}. However, these methods seem most effective in the low-rate and high-distortion regimes, and do not scale well into the high-rate, low-distortion regime which is the most desirable for the compression payload of a future mission. Indeed, as we show in this paper, CCSDS-123.0-B-2 remains substantially better than current deep learning methods. This calls for a significant departure in the design philosophy from the RGB-inspired autoencoders.

In this paper, we present a novel design for a lossless and near-lossless predictive coding neural network, called LineRWKV. The design departs from existing works in multiple significant ways. First, it follows a predictive coding approach, where the neural network predicts a pixel value from a causal context and only the prediction residual is entropy-coded. This is in contrast with the typical autoencoder design (more reminiscent of transform coding), where the bottleneck may be hard to scale for low-distortion, high-rate regimes. 
Importantly, to the best of our knowledge, current designs of hyperspectral compression techniques based on autoencoders only target lossy, but not lossless compression; in contrast, the proposed method can perform both lossless and lossy compression.
Moreover, we propose a neural network architecture that can work recursively in the along-track direction. This limits memory usage, as it is only tied to one image line with all its spectral channels, while exploiting a memory of past lines and enabling continuous pushbroom operation. Critically, we use a novel hybrid attentive-recursive operation \cite{peng2023rwkv} which avoids pitfalls of older recurrent neural networks (RNNs), enabling parallelized training, and having the representational power of Transformers \cite{vaswani2017attention}, with the advantage of linear complexity instead of quadratic. While \cite{peng2023rwkv} has been proposed for natural language processing tasks, we are the first to adapt this architecture to image processing; we note that, besides compression, the same idea could be employed for other tasks such as image analysis. We conduct extensive experiments with data from the EnMAP and PRISMA satellites to show that, for the first time for a deep-learning method, we can significantly outperform CCSDS-123.0-B-2 in rate-distortion performance. Finally, we conduct tests on a 7W embedded system to validate promising throughput results.

A preliminary version of this work \cite{valsesia2024hybrid} introduced the general idea, while this paper significantly expands the methodology and experimental results.

\section{Background}
\label{sec:bkg}
\subsection{Compression onboard of spacecrafts}
Onboard hyperspectral image compression faces a challenging balance between rate-distortion performance and computational complexity. Over the years, CCSDS standards have gained popularity due to their low computational demands, enabling high throughput on dedicated FPGAs \cite{santos2015multispectral}, and good rate-distortion performance. In particular, CCSDS 122.0-B-2 \cite{ccsds122} together with CCSDS 122.1-B-1 \cite{ccsds1221} employ a transform coding approach to losseless and lossy compression, where a 2D discrete wavelet transform is coupled with a spectral transform. The more recent CCSDS 123.0-B-2 \cite{ccsds123} follows instead a predictive coding approach to lossless and near-lossless (i.e., bounded error) compression with a spatial-spectral predictor based on an adaptive filter. CCSDS 123.0-B-2 is regarded as the state-of-the-art for onboard hyperspectral image compression and, to the best of our knowledge, remains unchallenged by current deep-learning methods. Finally, it worth mentioning that Regression Walevelet Analysis (RWA) \cite{amrani2016regression,alvarez2019regression} combines a spectral transform with 2D JPEG2000 to achieve compression performance competitive with CCSDS 123.0-B-2.

\subsection{Image compression with deep learning}
Most of the works in the literature of image compression with deep learning stem from the computer vision community, focusing on 8-bit RGB or grayscale images \cite{balle2016end,zhou2018variational,balle2018variational,balle2020nonlinear,jiang2023mlic,mentzer2019practical}. Few works \cite{alves2021reduced,chong2021end,dua2021convolution,la2022hyperspectral,mijares2023scalable} have addressed the topic of hyperspectral images, especially while keeping complexity in mind for possible onboard usage. 

The general approach mainly follows the early work of Ball\'e et al. \cite{balle2016end}, in the design of autoencoder neural networks where an encoder subnetwork creates a latent representation of the input image in a low-dimensional space which is then quantized and entropy-coded. A decoder subnetwork maps the latent representation back to the image space. These networks are trained end-to-end with rate-distortion objectives, balancing, according to the desired tradeoff, the reconstruction error and the entropy of the latent representation. Recent advancements for RGB images focused on improving encoder architectures and context models for the entropy coder \cite{jiang2023mlic}. Regarding hyperspectral images, the main challenge lies in the extremely high memory requirements to capture 3D spatial-spectral features for a compressed latent representation. In a recent work targeting onboard compression, Verd\'u et al. \cite{mijares2023scalable} adopt a channel clusterization strategy as well as novel normalization strategies needed for images with high dynamic range. While the method works well at low rates, it is difficult to scale to high-rate, high-quality regimes without compromising efficiency. In fact, the autoencoding approach tends to struggle when very low distortion is desired, if the network capacity in terms of number of features, latent space size and total number of parameters is not adequately scaled.

Finally, it is worth mentioning that some autoregressive approaches \cite{toderici2015variable,islam2021image} have been studied. They generally rely on modeling the image as a sequence of all pixels, and use a causal context for prediction within the mechanism of either causal convolution, RNNs or Transformers. As detailed in the next section, until recently, each of these approaches had a critical limitation,  consisting in either limited representational power, or inefficient training and encoding, or high complexity.
We also remark that the proposed method is not purely autoregressive on all image pixels, since the across-track dimension can be effectively processed in parallel, as detailed in Sec. \ref{sec:train_v_test}.

\subsection{Deep sequence processing}
\label{sec:bkg_seq}
Since the proposed method is based on a causal prediction approach, it is worth reviewing the main ways in which deep neural networks are used to process sequences and their respective tradeoffs. For simplicity we are going to discuss processing of a 1D sequence. 

First, a simple design would be causal convolution, such as that of PixelCNN \cite{van2016conditional}, where the convolution kernel is masked so that the receptive field only expands in the past samples. While this approach enjoys fairly efficient implementations, it has limited representational power as it is not an input-dependent operation (like Attention is), and the size of the receptive field, i.e., how far past samples affect the current prediction, might be limited.

An alternative approach would be the use of RNNs, such as the LSTM \cite{hochreiter1997long}, to have ideally infinite memory. This approach is limited by the inefficiency of training which requires serial computations over the entire sequence length, so that it does not scale to the large datasets required to train even moderately complex models.

Transformers \cite{vaswani2017attention} are currently the state of the art for sequence processing thanks to the high representational power of the input-dependent attention operation, which is capable of creating operations that are adaptive to the sequence under processing. Unlike RNNs, Transformers can also be parallelized during training. However, they suffer from quadratic complexity in the sequence length and require to keep the entire sequence and its features in memory, resulting in high memory and computational cost. This renders them infeasible to process an image in an autoregressive pixelwise fashion.

Very recently, the natural language processing literature has focused its attention towards sequence processing with hybrid designs between RNNs and Transformers which simultaneously have linear complexity, admit parallel training and can be written in a recursive manner for low-memory inference \cite{peng2023rwkv,gu2023mamba}. RWKV \cite{peng2023rwkv} is a recent model with such properties, which is at the basis of the proposed work, as detailed in Sec. \ref{sec:method}.

\section{Proposed method}
\label{sec:method}

\begin{figure*}
    \centering
    \includegraphics[width=0.85\textwidth]{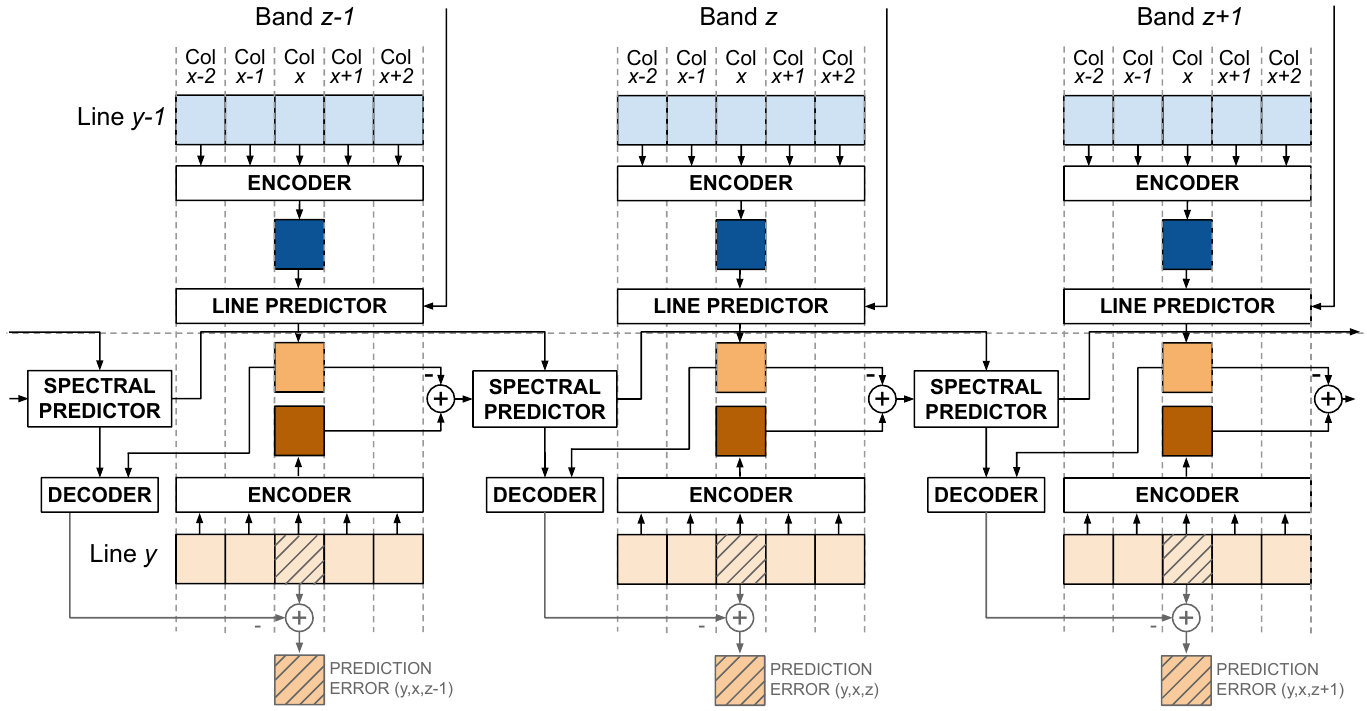}
    \caption{Overview of LineRWKV for $z>1$ and $y>0$. Encoder, decoder, line and spectral predictors are neural networks trained end-to-end to minimize the prediction error for all pixels. Each square represents a pixel in the corresponding spatial-spectral position, as it is processed by the neural modules. Rounded prediction errors are then entropy-coded with a standard code.}
    \label{fig:architecture}
\end{figure*}

\begin{figure*}
    \centering
    \includegraphics[width=0.7\textwidth]{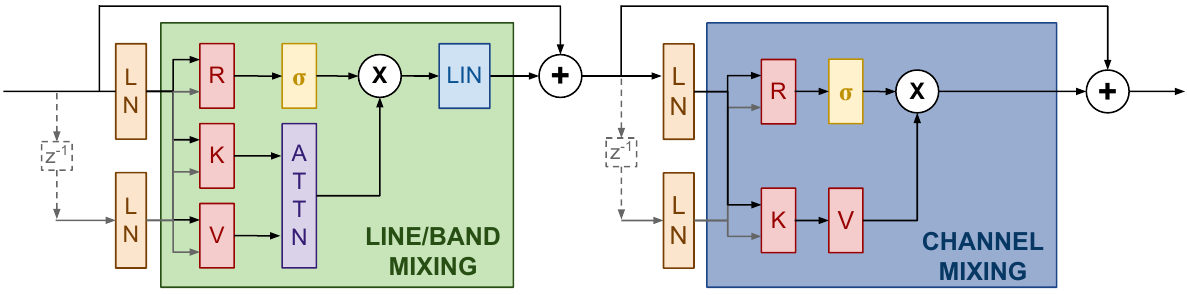}
    \caption{Line and spectral predictors are made of the repetition of line/band mixers and channel mixers according the depicted architecture. $z^{-1}$ denotes unit delay in the sequence dimension (lines or bands). LN denotes LayerNorm and LIN denotes linear projection of features.}
    \label{fig:line_predictor}
\end{figure*}

\subsection{Overview}
In this section, we present the full design of the proposed hyperspectral compression method. This comprises a prequantizer, followed by a deep learning based predictor, called LineRWKV, and an entropy coder to code the prediction residuals. For prequantization and entropy coding we employ traditional techniques, therefore in this section we focus on the description of the LineRWKV module. For clarity of explanation, we will describe its behavior in inference mode for the compression stage. Discrepancies with respect to the training process are discussed in Sec. \ref{sec:train_v_test}.

A high-level overview of LineRWKV is presented in Fig. \ref{fig:architecture}. We denote sampling of the along-track direction of the satellite with the term ``lines'' and variable $y=0,1,2,\dots$; the across-track direction is denoted as columns and variable $x=0,1,2,\dots,N_x-1$; the spectral direction is denoted as $z=0,1,2,\dots,N_z-1$. LineRWKV does not pose any restrictions on the extent of the three dimensions, which may be different between training and testing. In particular, the maximum number of columns and bands is dictated by available memory, while the number of lines can be infinite, supporting continuous pushbroom acquisition.

The general principle followed by LineRWKV is that of predictive coding. The architecture is a model that, based on a causal context of past spatial and spectral pixel values, predicts the next pixel value. In particular, as it will be explained in the following, the prediction for pixel $I_{y=Y,x=X,z=Z}$ depends on all pixels with $y\leq Y$, $z<Z$, $\forall x$. We note that the context employed by the encoder for the spatial and spectral predictor is slightly different from conventional designs. Specifically, the predictor employs all lines above the current pixel, and all previous bands for the current pixel, but it does not use information from the previous pixels on the same line or future bands in past lines. This is due to the LineRWKV design, which employs separate blocks in the spectral, along-track and across-track dimensions. This choice is slightly suboptimal; it is similar in principle to the ``narrow'' mode of the CCSDS-123.0-B-2 predictor, which incurs a small loss of compression efficiency \cite{valsesia2019high}, but it is necessary in order to avoid a much more complex structure of the neural predictor.
The predicted value is rounded and subtracted from the original pixel value to compute the prediction error. Finally, the prediction error is losslessly encoded with any entropy encoder.  At any time during inference, the last two lines of the image with all their spectral channels (i.e., a tensor of size $2 \times N_x \times N_z$) are supposed to be under processing to predict the next line, which is also available for the computation of the residual.

Lossy compression can be achieved in two ways: in-loop quantization of prediction errors or image prequantization. In-loop quantization requires to reconstruct the pixel value from the quantized residual in order to complete prediction for the next values, introducing serial data dependencies. While theoretically superior, this leads to inefficient implementations. On the other hand, prequantization consists in quantizing the input image and then using lossless compression. This avoids data dependencies at the compressor since there is no discrepancy between the original value and the value to be used for prediction, so it leads to higher compression throughput. At the high rates desirable for usage in real missions, the rate-distortion penalty with respect to in-loop quantization is minimal \cite{valsesia2019high}, so this is the preferred option for this work.

The LineRWKV model has a modular design composed of the following blocks: i) an \textit{encoder} to map each pixel in a line and given band into a feature space by exploiting across-track spatial correlation; ii) a \textit{line predictor} that predicts the features of the co-located (same $x,z$) pixel in the next line; iii) a \textit{spectral predictor} that predicts the features of the co-located (same $x,y$) pixel in the next band; iv) a \textit{decoder} that estimates a pixel value based on its features. The following sections explain each of the blocks in detail. In the following, we suppose that we are predicting the values for line $y$ from current line $y-1$ and past line $y-2$, just as depicted in Fig. \ref{fig:architecture}.

\subsection{Encoder}\label{sec:method_encoder}

The goal of the encoder is to capture the correlation that exists across image columns and encode it into a feature space for further processing. The encoder function is thus a 1D operation that is shared for all bands and for all lines. A simple design for the encoder is a sequence of blocks composed of 1D convolution, layer normalization \cite{ba2016layer}, and non-linear activation. This design should be modulated according to the expected complexity of the across-track correlation patterns, possibly considering larger receptive fields and attention operations, if complexity allows.
For our experiments, we choose the simple convolutional blocks and we denote the number of such blocks with $N_\text{enc}$. The encoding of the input line will result in a feature vector for each spatial-spectral pixel (i.e., a tensor of size $1 \times N_x \times N_z \times F$).

\subsection{Line Predictor}
\label{sec:method_line}

The line predictor is the core operation of LineRWKV, which enables the recursion over the image lines, constraining memory usage and enabling continuous operation. The line predictor should exploit the features of a number of past lines in order to predict the next line. As discussed in Sec. \ref{sec:bkg_seq}, it is desirable to have an attention-based operation such as the scaled dot-product attention of the Transformer to process the sequence of lines, both to parallelize training and to exploit the power of attention to adapt to self-similar patterns. However, Transformers require keeping several lines and their features in memory, which is prohibitive, and have quadratic computational cost in the number of lines, which is also prohibitive. On the other hand, training a traditional RNN on large-scales would be extremely slow and would not scale well. Our line predictor is therefore based on the RWKV neural network \cite{peng2023rwkv} for sequence processing, and its recurrent implementation is used for inference.

More in detail, each of the $N_xN_z$ pixels in the current line, represented as a feature vector by the across-track encoder, is processed in parallel by the line predictor. The RWKV-based line predictor is composed of the repetition of two fundamental blocks: line mixing and channel mixing, both depicted in Fig. \ref{fig:line_predictor}. Overall, $N_\text{lp}$ line and channel mixing blocks are used in our architecture.  For brevity of notation, we denote the feature vector of a pixel input to the line mixing operation as $\mathbf{u}_{y-1} \in \mathbb{R}^F$ and its corresponding output as $\mathbf{o}_{y-1} \in \mathbb{R}^F$.
In particular, line mixing performs the following operations:
\begin{align}
    \mathbf{r}_{y-1} &= \mathbf{W}_r (\mu_r\mathbf{u}_{y-1} + (1-\mu_r)\mathbf{u}_{y-2})\\
    \mathbf{k}_{y-1} &= \mathbf{W}_k (\mu_k\mathbf{u}_{y-1} + (1-\mu_k)\mathbf{u}_{y-2})\\
    \mathbf{v}_{y-1} &= \mathbf{W}_v (\mu_v\mathbf{u}_{y-1} + (1-\mu_v)\mathbf{u}_{y-2})
\end{align}
\begin{align}
    \mathbf{a}_{y-1} &= e^{-\bm{\alpha}} \odot \mathbf{a}_{y-2} + e^{\mathbf{k}_{y-1}} \odot \mathbf{v}_{y-1}\\
    \mathbf{b}_{y-1} &= e^{-\bm{\alpha}} \odot \mathbf{b}_{y-2} + e^{\mathbf{k}_{y-1}}\\
    \mathbf{p}_{y-1} &= \frac{\mathbf{a}_{y-2} + e^{\bm{\beta}+\mathbf{k}_{y-1}} \odot \mathbf{v}_{y-1}}{\mathbf{b}_{y-2}+e^{\bm{\beta}+\mathbf{k}_{y-1}}}\\
    \mathbf{o}_{y-1} &= \mathbf{W}_o (\sigma(\mathbf{r}_{y-1}) \odot \mathbf{p}_{y-1})
\end{align}
where $\mathbf{W}_*$, $\bm{\alpha}$, $\bm{\beta}$ are trainable parameters, $\mu_*$ some hyperparameters and $\sigma$ the sigmoid function. The symbol $\odot$ denotes elementwise product and exponentiation to a vector is intended elementwise. The states in the recursion are initialized as $\mathbf{a}_{0}=\mathbf{b}_0=\mathbf{0}$.

With some abuse of notation, also denoting with $\mathbf{u}_{y-1} \in \mathbb{R}^F$ the input to the channel mixing module and its corresponding output as $\mathbf{o}_{y-1} \in \mathbb{R}^F$, the module performs the following operations:
\begin{align}
    \mathbf{r}_{y-1} &= \mathbf{W}'_r (\mu'_r\mathbf{u}_{y-1} + (1-\mu'_r)\mathbf{u}_{y-2})\\
    \mathbf{k}_{y-1} &= \mathbf{W}'_k (\mu'_k\mathbf{u}_{y-1} + (1-\mu'_k)\mathbf{u}_{y-2})\\
    \mathbf{o}_{y-1} &= \sigma(\mathbf{r}_{y-1}) \odot \left( \mathbf{W}'_v \text{max}(\mathbf{k}_{y-1},0)^2 \right)
\end{align}
where $\mathbf{W}'_*$, are trainable parameters, $\mu'_*$ some hyperparameters, and $\sigma$ the sigmoid function.

It can be noticed that the recursion requires a limited amount of memory to be implemented. In particular, besides the features of the current line, the only feature vectors that need to be stored are the states $\mathbf{a}_{y-2}$ and $\mathbf{b}_{y-2}$ and features of the previous line $\mathbf{u}_{y-2}$. This needs to be done for each layer of line and channel mixing that is used. However, this operation allows to keep a memory of past lines and use the attention mechanism for the prediction of the next line and is significantly advantageous in terms of memory consumption compared to keeping the features of a large number of lines. We refer the reader to \cite{peng2023rwkv} for details about a numerically-stable implementation.

\subsection{Spectral Predictor}
\label{sec:method_spectral}

The output $\mathbf{o}_{y-1}$ of the line predictor is a feature vector for each spatial-spectral pixel that can be thought to be a prediction of the features of line $y$. While this captures the spatial correlation in the across- and along-track directions, the model has not yet exploited correlation in the spectral dimension.
We propose to first compute a feature-domain spatial prediction residual and then model its spectral correlation. In particular, the pixels in line $y$ are encoded with the encoder in Sec. \ref{sec:method_encoder} and the difference between them and the output of the line predictor applied to line $y-1$ forms the spectral sequence $\lbrace \bm{\Delta}_z \rbrace_{z=1}^{N_z}$. Notice that $\bm{\Delta}_{y,x,z}$ requires pixel $I_{y,x,z}$ for its computation. Therefore, this needs to have been already decoded and its availability is ensured by treating the first band as a special case, discussed in Sec. \ref{sec:method_special}. 

A causal model over the $\bm{\Delta}_z$ sequence is needed to obtain features describing the pixel to be predicted. For this spectral prediction model, we propose to also use RWKV blocks. However, for a lossless (or prequantized) compressor, in order to ensure high throughput, the parallel implementation typically reserved for training is used rather than the recurrent implementation described in Sec. \ref{sec:method_line}. This consists in a sequence of $N_\text{sp}$ band mixing and channel mixing blocks, with the same architecture previously depicted in Fig. \ref{fig:line_predictor}. The parallel implementation keeps the features for all the bands in memory and computes\footnote{With some abuse of notation we reuse symbols from the line predictor, but they represent different activations and weights.}:
\begin{align}
    \mathbf{r}_{z-1} &= \mathbf{W}_r (\mu_r\bm{\Delta}_{z-1} + (1-\mu_r)\bm{\Delta}_{z-2})\\
    \mathbf{k}_{z-1} &= \mathbf{W}_k (\mu_k\bm{\Delta}_{z-1} + (1-\mu_k)\bm{\Delta}_{z-2})\\
    \mathbf{v}_{z-1} &= \mathbf{W}_v (\mu_v\bm{\Delta}_{z-1} + (1-\mu_v)\bm{\Delta}_{z-2})\\
    \mathbf{p}_{z-1} &= \frac{\sum_{i=1}^{z-2} e^{-(z-1-i)\bm{\alpha}+\mathbf{k}_i} \odot \mathbf{v}_i+e^{\bm{\beta}+\mathbf{k}_{z-1}} \odot \mathbf{v}_{z-1}}{\sum_{i=1}^{z-2} e^{-(z-1-i)\bm{\alpha}+\mathbf{k}_i}+e^{\bm{\beta}+\mathbf{k}_{z-1}}} \\
    \mathbf{o}_{z-1} &= \mathbf{W}_o (\sigma(\mathbf{r}_{z-1}) \odot \mathbf{p}_{z-1})
\end{align}
for the band mixing block, and:
\begin{align}
    \mathbf{r}_{z-1} &= \mathbf{W}'_r (\mu'_r\mathbf{u}_{z-1} + (1-\mu'_r)\mathbf{u}_{z-2})\\
    \mathbf{k}_{z-1} &= \mathbf{W}'_k (\mu'_k\mathbf{u}_{z-1} + (1-\mu'_k)\mathbf{u}_{z-2})\\
    \mathbf{o}_{z-1} &= \sigma(\mathbf{r}_{z-1}) \odot \left( \mathbf{W}'_v \text{max}(\mathbf{k}_{z-1},0)^2 \right)
\end{align}
for the channel mixing block.
The output of the spectral predictor is a feature vector for each pixel which should ideally represent the spatial-spectral residual in a feature domain. Also notice that spectral prediction is performed on the difference sequence $\bm{\Delta}$ rather than on the outputs of the line predictors as this allows to use the information of the pixel in the same line to be encoded but from previous bands, which is typically the highest source of correlation.

\subsection{Decoder}
\label{sec:method_decoder}

A decoder neural network produces a prediction of the raw pixel value from the feature-domain representation of the spatial prediction and spatial-spectral error. This is obtained by feeding their sum to a sequence of $N_\text{dec}$ blocks composed of $1\times1$ convolution, LayerNorm and non-linear activation. 

The predicted pixel value is then denormalized, rounded to the nearest integer and the difference with respect to the original pixel value forms the prediction error, which is entropy-coded with a suitable technique. Notice that this approach differs somewhat from works addressing lossless compression of natural images with deep learning \cite{mentzer2019practical}. In those works, it is more typical to let the neural network produce the prediction as a probability distribution over the set of possible symbols (typically 256 for 8-bit images). However, the large number of symbols (e.g., $2^{16}$) of satellite images poses efficiency challenges in terms of computation and memory requirements and incurs in the ``softmax bottleneck'' issue due to the number of classes exceeding the number of features \cite{yang2018breaking}. The proposed floating-point regression with rounding solves these issues but incurs in subtle numerical conditions that must be managed to ensure decodability. Whenever there is a mismatch in the hardware-software stack between the compressor and the decompressor, the prediction will only be accurate down to numerical precision, i.e., about 7 significant digits on the normalized value, for computations in FP32. This means that a numerical perturbation might cause the predicted value to cross the rounding threshold, causing a decoding error. In order to avoid this, whenever the prediction is closer to the rounding threshold than $10^{-3}$ (in unnormalized integer digital numbers), we signal the side of the threshold as extra information. This choice has been experimentally verified to ensure correct decoding on the entire HySpecNet-11k test set. The side information is entropy-coded and in our tests incurred a small penalty of 0.02 bpppc, already included in all experimental results.

\subsection{Special cases}
\label{sec:method_special}
In order to ensure causality of the model, the first line for all bands and the entire first band need to be compressed with a separate method. For the first line, we use a simple DPCM encoder where the prediction error is computed as:
\begin{align*}
    E_{0,x,z} = \begin{cases}
     I_{0,0,0}, & \text{for } x=0, z=0\\
     I_{0,x,z} - I_{0,x-1,0}, & \text{for } x>0, z=0\\
     I_{0,x,z} - I_{0,x,z-1} & \text{for } x\geq0, z>0.
    \end{cases}
\end{align*}
For the first band, we employ spatial prediction only, by feeding the output of the line predictor to a dedicated decoder neural network with architecture comparable to that described in Sec. \ref{sec:method_decoder}; the prediction error is then entropy coded.

\subsection{Training vs. Inference}
\label{sec:train_v_test}
The description in the previous section follows the inference behavior. A few important differences are present for training in order to ensure scalability when using large datasets. First, the recurrent implementation of the line predictor is replaced with the parallel implementation already described in Sec. \ref{sec:method_spectral}. This improves efficiency as it enables processing several lines at the same time, and, in fact, it is one of the main advantages over traditional recurrent neural networks. Moreover, we remark an interesting property of the overall design, which is useful for efficient training: all operations performed after the line encoder are columnwise independent. In fact, the predictors work on lines and bands but do not mix columns, and the use of $1\times1$ convolution in the decoder also keeps this property. This is useful to reduce memory consumption during training by subsampling random subsets of columns after encoding, if desired.

LineRWKV is trained by minimizing the $\ell_1$ loss between the predicted and true pixel values. Notice that this means that we only optimize for lossless compression, while never accounting for quantization during training. Nevertheless, lossy rate distortion tests still show excellent performance, as reported in Sec. \ref{sec:results}.

\section{Experimental results}
\label{sec:results}

In this section, we analyze the performance of LineRWKV in terms of compression efficiency with respect to state-of-the-art approaches to hyperspectral image compression. Moreover, we show how LineRWKV behaves when a different satellite is targeted compared to its original training. Finally, we report throughput and memory usage on a 7W low-power device as a proof of concept for potential onboard implementation on embedded devices. 

\begin{table}[t]
\centering
\caption{LineRWKV architecture configurations.}
\label{tab:linerwkv_configs}
\setlength{\tabcolsep}{4pt}
\begin{tabular}{cccccccc}
\textbf{Model Size} & \textbf{Params} & $N_\text{enc}$ & $N_\text{lp}$ & $N_\text{sp}$ & $N_\text{dec}$ & $F$ & \textbf{FLOPS/sample} \\
\hline
\hline
XS                  & 30k             & 1               & 2              & 2              & 1               & 32  &  120k     \\
S                   & 135k            & 2               & 2              & 2              & 2               & 64  &  508k     \\
M                   & 286k            & 4               & 4              & 4              & 4               & 64  &  1M     \\
L                   & 900k            & 4               & 6              & 6              & 4               & 96  &  3.2M    \\
\hline
\end{tabular}
\end{table}

\subsection{Experimental setting}
Our main experimental results are based on training and testing LineRWKV on the recently introduced HySpecNet-11k dataset \cite{fuchs2023hyspecnet}. This is the largest curated dataset of hyperspectral images currently available, composed of 11,483 non-overlapping patches of size $128 \times 128 \times 224$ acquired by the EnMAP satellite with a ground sampling distance of 30m. The authors provide standard train-test splits for benchmarking compression algorithms. In particular, we use the ``hard'' split where patches in the test set belong to entirely separate tiles with respect to the training patches. Preprocessing discards some bands, resulting in a total of 202, and clips values between 0 and 10000.

We consider several baseline methods for hyperspectral image compression, focusing on low-complexity methods. In particular, we choose the state-of-the-art CCSDS standard for onboard lossless and near-lossless hyperspectral predictive compression, i.e., CCSDS-123.0-B-2 \cite{ccsds123}, and RWA \cite{amrani2016regression} for an alternative state-of-the-art approach to lossless compression based on transforms. As for state-of-the-art deep learning approaches, we evaluate the 1D-CAE \cite{kuester20211d}, SSCNet \cite{la2022hyperspectral} methods as well as the recent method by Verd\'u et al. \cite{mijares2023scalable}; we remark that these methods only perform lossy compression. All methods have been retrained, using the reference implementation by the HySpecNet-11k authors for the former two and the authors' code for the latter.  
We use the sample-adaptive Golomb encoder defined by the CCSDS-123.0-B-2 standard \cite{ccsds123} as entropy coder for the positive-mapped prediction residuals, both for our method and the CCSDS method to ensure a fair comparison; it is obvious that the performance could be improved employing an arithmetic coder, especially at rates below 2 bpppc, and as a matter of fact, the Golomb coder is not capable of producing rates below 1 bpppc. The other methods use the same entropy encoder as proposed in their respective works. The CCSDS predictor is used in its full mode with wide neighbor-oriented local sums, and 3 prediction bands, a configuration which typically provides the best tradeoff of compression ratio and complexity. 

Several configurations for LineRWKV have been tested in order to validate its scaling potential as function of number of encoder, decoder, predictors layers and number of features. Table \ref{tab:linerwkv_configs} reports the details for four configurations of choice, ranging from extra small (XS) to large (L).
Training LineRWKV has been done over four Nvidia A100 40GB GPUs. In order to limit training memory requirements and speed up convergence, a random subset of 16 contiguous bands has been used in the initial training phases, followed by finetuning with all the available bands and a smaller batch size. For the M configuration, the initial batch size was 8 without any column sampling; in the finetuning phase will all bands, the batch size is 4 with a subsampling of 16 columns. Learning rate was linearly decreased from $10^{-4}$ to $10^{-6}$ over 4000 epochs. We remark that memory requirements during training are large due to the desire to process as much of the hyperspectral cube as possible concurrently with a large enough batch size. This is not reflected in the inference phase, which is memory-efficient as shown in Sec. \ref{sec:exp_perf}.

\begin{table}[t]
\centering
\caption{Lossless rate (bpppc) on HySpecNet-11k hard test set.}
\label{tab:linerwkv_lossless}
\setlength{\tabcolsep}{1.5pt}
\begin{tabular}{cccccc}
\textbf{Model size} & \textbf{LineRWKV} & \textbf{CCSDS} \cite{ccsds123} & \textbf{RWA} \cite{amrani2016regression} &  \textbf{Diff. CCSDS} & \textbf{Diff. RWA} \\
\hline
\hline
XS &  \textbf{5.647} &  5.801 & 5.772  &  -0.154 & -0.125                 \\
S  &  \textbf{5.521} &  5.801 & 5.772  &  -0.280 & -0.251               \\
M  &  \textbf{5.510} &  5.801 & 5.772  &  -0.291 & -0.262               \\
L  &  \textbf{5.370} &  5.801 & 5.772  &  -0.431 & -0.402              \\
\hline
\end{tabular}
\end{table}

\begin{figure}[t]
    \centering
    \includegraphics[width=0.95\columnwidth]{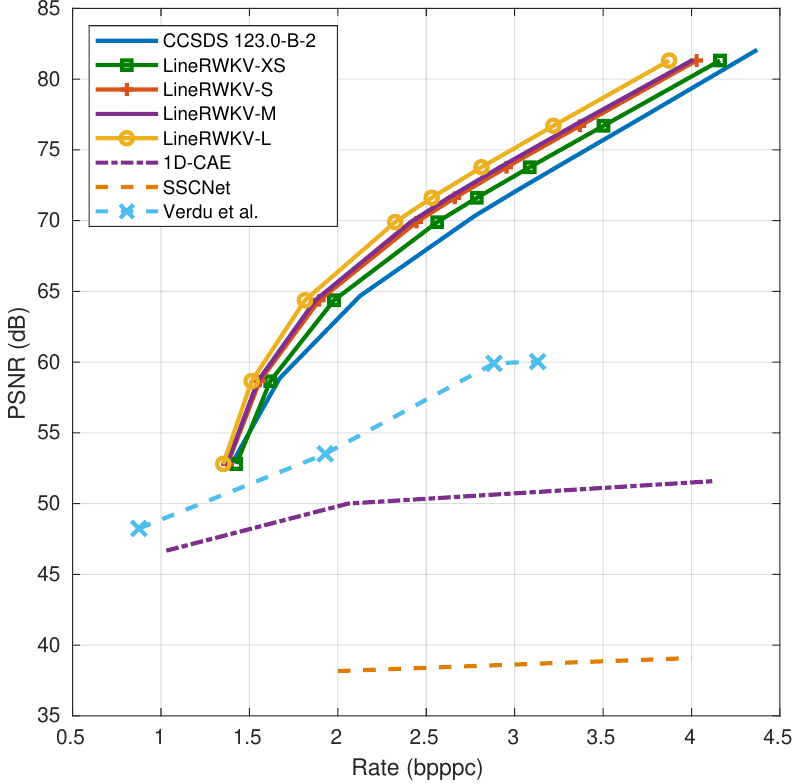}
    \caption{Rate-PSNR performance comparison on HySpecNet-11k hard test set.}
    \label{fig:ratedist}
\end{figure}

\begin{figure}[t]
    \centering
    \includegraphics[width=0.32\columnwidth]{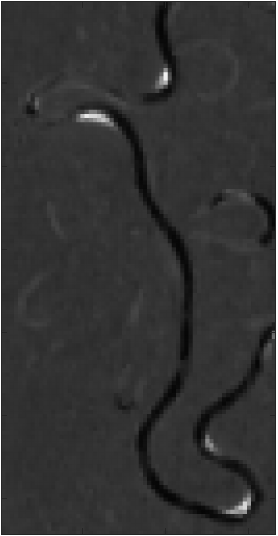}
    \includegraphics[width=0.32\columnwidth]{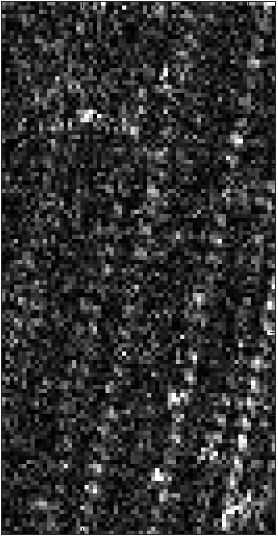}
    \includegraphics[width=0.32\columnwidth]{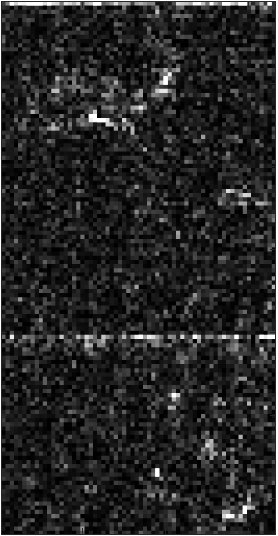}
    \caption{Left to right: band 187 of a test image (colormap $[0,4000]$, positive-mapped prediction residuals (colormap $[0,100]$) for CCSDS-123.0-B-2 and LineRWKV-XS.}
    \label{fig:viz}
\end{figure}

\begin{figure}[]
    \centering
    \includegraphics[width=0.95\columnwidth]{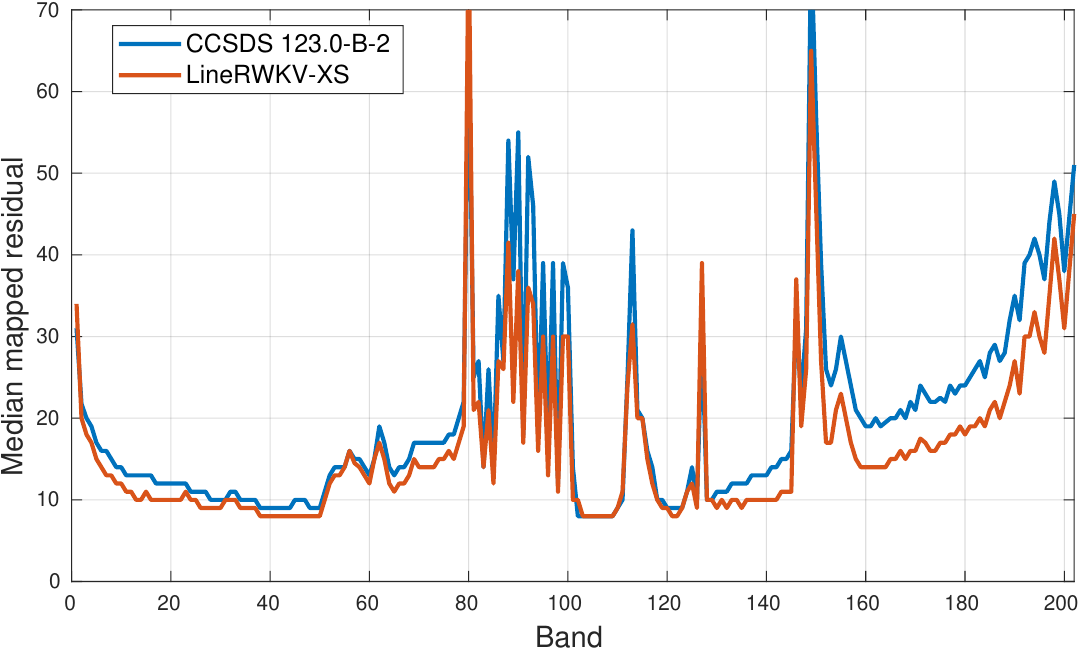}
    \caption{Positive-mapped prediction residual for CCSDS-123.0-B-2 and LineRWKV-XS. Median, over the entire HySpecNet-11k hard test set, of spatial medians.}
    \label{fig:res_per_band}
\end{figure}

\begin{figure*}[]
    \centering
    \includegraphics[width=0.3\textwidth]{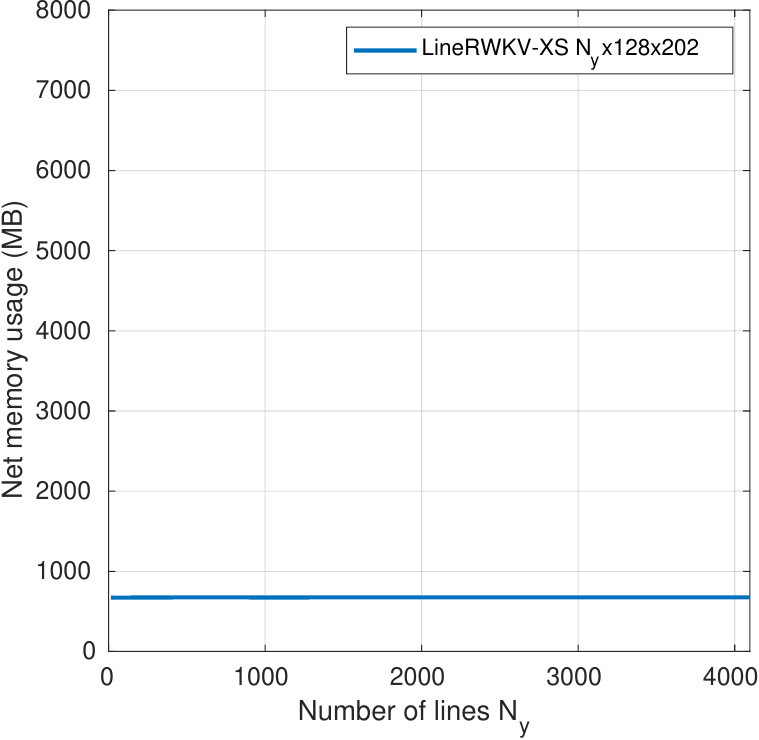} \quad
    \includegraphics[width=0.3\textwidth]{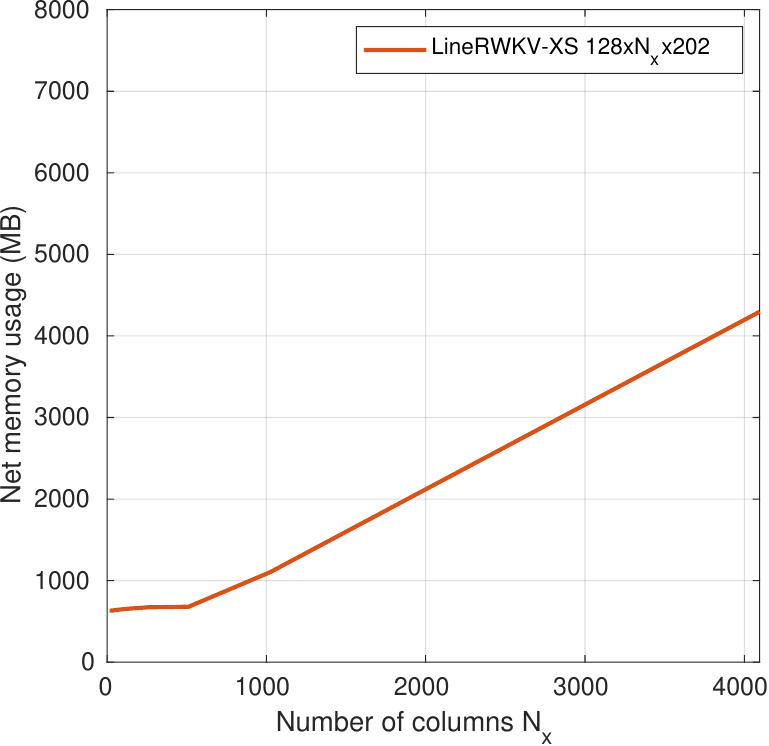} \quad
    \includegraphics[width=0.3\textwidth]{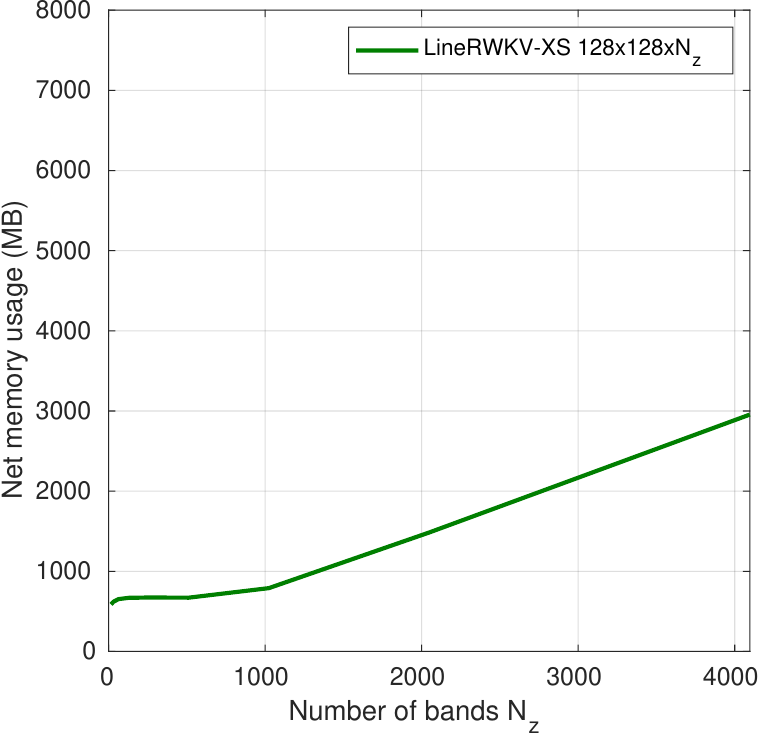}
    \caption{Scaling of LineRWKV-XS memory usage on the Nvidia Jetson Orin Nano 8GB as a function of number of lines, columns and bands.}
    \label{fig:jetson_scaling}
\end{figure*}

\subsection{Results on HySpecNet-11k}
Our main experiment assesses the rate-distortion performance of all methods when tested on the hard test split of HySpecNet-11k. We are mostly interested in the high-quality and high-rate regimes (typically $\geq2$ bpppc) as these are the most relevant for space missions.

Fig. \ref{fig:ratedist} shows the average PSNR obtained over the entire test set as a function of the rate for all the methods under study. It can be noticed that existing deep learning approaches are significantly outperformed by the CCSDS standard at high rates. This is due to the current focus of the literature on hyperspectral image compression towards very low rates, which finds its motivation in complex challenges towards scaling neural network designs to also work well at high rates while maintaining acceptable complexity.
As a sanity check, we remark that the reported values on the hard split are consistent with published results on the easy split by the HySpecNet-11k authors. Concerning LineRWKV, we can notice that it outperforms CCSDS-123.0-B-2, even in its extra small configuration. Indeed, we notice a clear performance scaling, with increasing rate-distortion gains over CCSDS as model complexity is increased. This is also evident on the rates achieved for lossless compression, which are reported in Table \ref{tab:linerwkv_lossless}. The XS configuration reports an already substantial 0.154 bpppc gain, while the L configuration achieves a massive 0.431 bpppc gain. Although the L configuration is too complex for practical onboard usage, it serves as a validation that LineRWKV has potential scaling and is not significantly bottlenecked in its design. In the same table, we also see that RWA \cite{amrani2016regression} is slightly better than CCSDS 123.0-B-2 at lossless compression, but it is still outperformed by the proposed LineRWKV.

Fig. \ref{fig:viz} shows a qualitative comparison of the positive-mapped prediction residuals of LineRWKV-XS and CCSDS-123.0-B-2 for a band in a test image. It can be noticed that some complex correlation patterns are discovered by the neural network and removed, leading to smaller residuals. On the other hand, Fig. \ref{fig:res_per_band} reports the median positive-mapped prediction residual as a function of band index for the entire HySpecNet-11k hard test set. We can see that LineRWKV-XS provides smaller residuals across basically all the 202 bands, with the largest gains observed for bands 130 to 202.

As a term of comparison regarding complexity, the work by Verd\'u et al. \cite{mijares2023scalable} reports a complexity of about 17k FLOPs/sample compared to the 120k FLOPs/sample of LineRWKV-XS. However, we also tested a variant of the method where we increased the number of hidden and latent features to approximately match 120k FLOPs/sample. This is the point with rate equal to 3.130 bpppc reported in Fig. \ref{fig:ratedist}, which however does not seem to provide significant improvements, highlighting that the design has some bottleneck preventing it from reaching low distortion values.

\subsection{Transfer Learning to PRISMA}

\begin{table}[]
\centering
\caption{Transfer Learning to PRISMA - LineRWKV lossless rate difference vs. CCSDS-123.0-B-2.}
\label{tab:prisma}
\begin{tabular}{ccc}
\textbf{Model size} & \textbf{Zero-shot transfer} & \textbf{After finetuning} \\
\hline
\hline
M          &       +0.053             & -0.160      \\    
\hline
\end{tabular}
\end{table}

We are interested in studying how well a LineRWKV model trained on a dataset from a specific satellite behaves when it is applied to a different satellite. This is important since it is likely that a large-scale dataset of images representative of a new satellite is not available before launch. However, it is also conceivable that a certain number of images can be losslessly transmitted with a suboptimal compressor in the pre-operational phase in order to finetune and update the compressor model itself. 

In order to perform this study, we collected 110 images of size $1000 \times 1000 \times 239$ from the PRISMA satellite \cite{cogliati2021prisma}. These images have been collected from all over the world to have a highly varied set of scenes, and the test locations are strictly disjoint from the train locations. For the transfer learning experiment, we used the 66 VNIR bands.
Table \ref{tab:prisma} reports the lossless rate reduction compared to CCSDS for the M configuration in two settings of interest: \textit{zero-shot transfer}, i.e., the model trained on HySpecNet-11k is used as is on PRISMA, and after \textit{finetuning} with 100 PRISMA images. It can be noticed that the domain gap between EnMAP and PRISMA affects compression performance in the zero-shot case which is now slightly outperformed by the CCSDS algorithm. However, finetuning with a modest number of images allows to recover superior performance.

\subsection{Performance on low-power hardware}
\label{sec:exp_perf}
In order to validate the low-complexity inference of LineRWKV, we test its performance on a low-power embedded system, namely the Nvidia Jetson Orin Nano. The platform has 8GB of shared CPU-GPU memory and a maximum power of 7W. We test the XS configuration in single-precision floating point (FP32). For the HySpecNet-11k images ($128 \times 128 \times 202$), the peak memory usage due to LineRWKV is 673 MB, which is a quite modest value in the realm of deep learning-based processing of hyperspectral images. The measured latency is 6.472 seconds, resulting in a throughput of 511,345 samples/sec. Fig. \ref{fig:jetson_scaling} shows how memory scales as a function of number of lines, columns and bands. First, we can notice that the recurrent property over the along-track direction allows constant memory usage regardless of the number of lines. For what concerns columns and bands, we can see that scaling is fairly linear, in contrast with the quadratic dependence in traditional Transformers, with a floor for small image size. Throughput is roughly constant around 500k samples/s for all image sizes except for the very small ones, which also confirms linear scaling of complexity.

It must be remarked that the current implementation is far from optimized. Indeed, several optimizations are possible that would substantially raise the throughput and might be worth investigating in future works. Few examples include model compilation for the target accelerator (e.g., via TensorRT), mixed-precision inference where parts of the model run integer-quantized or in floating point half precision. Our proposed design also does not exploit any spatial downsampling: if the line and spectral predictors could be run on a downsampled cube, throughput would be increased by a factor close to the downsampling factor. However, this necessitates careful decoder design not to reduce rate-distortion performance.

\subsection{Spectral Predictor Ablation}
In this section, we evaluate the choice of RWKV blocks as a spectral prediction mechanism against an alternative approach, presented in \cite{valsesia2024hybrid}. The alternative approach adopts causal convolution operations such that the receptive field expands only towards past bands. In particular, blocks are repeated to form the spectral predictor; each block is composed of 1D causal convolution with kernel length of 3 bands, Layer Normalization, GeLU non-linearity and an attention operation. This operation computes an attention mask from the output of the GeLU non-linearity with two size-1 convolutions interleaved by GeLU and a final sigmoid activation. The resulting mask is multiplied elementwise to the input of the attention operation and the result summed to it. Comparisons are made to have a similar number of trainable parameters with respect to the design with RWKV blocks.

Table \ref{tab:spectral_ablation} reports a comparison between the lossless rate and throughput on the Nvidia Jetson Orin Nano of some configurations of the proposed LineRWKV model with RWKV-based spectral predictor, and those of the model with causal convolution-based spectral predictor. We can see that the causal convolution method is faster but not as effective in terms of compression ratio.

\begin{table}[t]
\centering
\caption{Ablation of the design of spectral predictor.}
\label{tab:spectral_ablation}
\begin{tabular}{ccccc}
\multirow{2}{*}{\textbf{Model size}} & \multicolumn{2}{c}{\textbf{RWKV}}   & \multicolumn{2}{c}{\textbf{Causal conv.}} \\
                                     & \textbf{Rate} & \textbf{Throughput} & \textbf{Rate}    & \textbf{Throughput}    \\ \hline \hline
\textbf{XS}                          & 5.647         & 511,345             &  5.764         &  595,962                      \\
\textbf{M}                           & 5.510         & 173,078             & 5.593            & 334,476         \\ \hline     
\end{tabular}
\end{table}

\section{Conclusions}
\label{sec:conclusions}

We presented a novel design of a low-complexity neural network for compression of hyperspectral images with the goal of usage onboard of satellites. Exploiting a predictive coding approach and recent advancements in hybrid recurrent-attentive neural networks, we showed to be able to overcome the memory and complexity bottleneck of current designs. This leads to a significant improvement of rate-distortion performance, resulting in the first deep-learning approach capable of outperforming CCSDS-123.0-B-2. Future work will focus on refining the design as several approaches could be used to further increase the already promising throughput.

\bibliographystyle{IEEEtran}


\end{document}